\documentclass[letter]{aa}
\usepackage[varg]{txfonts}
\usepackage{graphicx}
\usepackage{epstopdf}
\usepackage{pdflscape}
\usepackage{booktabs}
\usepackage{txfonts}
\usepackage{gensymb}
\usepackage{color}
\usepackage{url}
\usepackage{multirow}
\usepackage{amsmath}
\usepackage{amstext}
\usepackage{natbib}
\usepackage{longtable}
\usepackage{float}
\usepackage[export]{adjustbox}

\begin{document}

\title{An Update of the Correlation between Polarimetric and Thermal Properties of Cometary Dust}

\author{Yuna G. Kwon\inst{\ref{inst1}\thanks{Alexander von Humboldt Postdoctoral Fellow}}~\and~Ludmilla Kolokolova\inst{\ref{inst2}} \and Jessica Agarwal\inst{\ref{inst1},\ref{inst3}} \and Johannes Markkanen\inst{\ref{inst1},\ref{inst3}}}

\institute{Institut f{\" u}r Geophysik und Extraterrestrische Physik, Technische Universit{\" a}t Braunschweig,  Mendelssohnstr. 3, 38106 Braunschweig, Germany\\
\email{y.kwon@tu-braunschweig.de}\label{inst1} 
\and Department of Astronomy, University of Maryland, College Park, 20742 MD, USA\label{inst2}
\and Max Planck Institute for Solar System Research, Justus-von-Liebig-Weg 3, 37077 G{\"o}ttingen, Germany\label{inst3}
}

\date{Received April 28, 2021 / Accepted May 26, 2021}

\abstract {Comets are conglomerates of ice and dust particles, the latter of which encode information on changes in the radiative and thermal environments to date. Dust displays distinctive scattered and thermal radiation in the visible and mid-infrared (MIR) wavelengths, respectively, based on its inherent characteristics.}
{We aim to identify a possible correlation between the properties of scattered and thermal radiation from dust and the principal dust characteristics responsible for this relationship, and therefrom to gain insights into comet evolution.}
{We use the NASA/PDS archival polarimetric data on cometary dust in the Red (0.62--0.73 $\mu$m) and K (2.00--2.39 $\mu$m) domains to leverage the relative excess of the polarisation degree of a comet to the average trend at the given phase angle ($P_{\rm excess}$) as a metric of the dust's scattered light characteristics. The flux excess of silicate emissions to the continuum around 10 $\mu$m ($F_{\rm Si}/F_{\rm cont}$) is adopted from previous studies as a metric of the dust's MIR feature.}
{The two observables --- $P_{\rm excess}$ and $F_{\rm Si}/F_{\rm cont}$ --- show a positive correlation when $P_{\rm excess}$ is measured in the K domain (Spearman's rank correlation coefficient $\rho$ = 0.71$^{\rm +0.10}_{\rm -0.19}$). No significant correlation was identified in the Red domain ($\rho$ = 0.13$^{\rm +0.16}_{\rm -0.15}$). The gas-rich comets have systematically weaker $F_{\rm Si}/F_{\rm cont}$ than the dust-rich ones, yet both groups retain the same overall tendency with different slope values. } 
{The observed positive correlation between the two metrics indicates that composition is a peripheral factor in characterising the dust's polarimetric and silicate emission properties. The systematic difference in $F_{\rm Si}/F_{\rm cont}$ for gas-rich versus dust-rich comets would rather correspond with the difference in their dust size distribution. Hence, our results suggest that the current MIR spectral models of cometary dust, which search for a minimum $\chi^{\rm 2}$ fit by considering various dust properties simultaneously, should prioritise the dust size and porosity over the composition. 
With light scattering being sensitive to different size scales in two wavebands, we expect the K-domain polarimetry to be sensitive to the properties of dust aggregates, such as size and porosity, which might have been influenced by evolutionary processes. On the other hand, the Red-domain polarimetry reflects the characteristics of sub-$\mu$m constituents in the aggregate. }


\titlerunning{Correlation between Polarimetric and Thermal Properties of Cometary Dust}

\authorrunning{Y. G. Kwon et al.}

\maketitle

\section{Introduction \label{sec:intro}}

Comets are among the most pristine leftovers from the nascent solar system and conglomerates of ice and dust. The dust properties can reflect changes in the radiative and thermal environment since formation. Observational constraints on these properties will thus bring valuable insights into the evolution of planetesimals in our planetary system.

Dust particles ejected from comets orbiting the Sun reflect sunlight mainly in the visible and near-infrared (NIR) wavelengths (0.5--2.5 $\mu$m), yielding a certain degree of linear polarisation \citep{Kiselev2015}. The remaining incident light is absorbed and re-emitted as thermal emission at longer wavelengths, particularly showing distinctive silicate emission features around 10 $\mu$m \citep{Hanner2004}. Both observables' behaviours are influenced by intrinsic microphysical (e.g., size distribution and internal structure) and compositional properties of the dust particles. Then, the following questions arise: Are these observables correlated? And if so, which dust properties contribute most to the correlation and what insights do they give us about the evolution of comets?

An earlier systematic study investigating a correlation between the polarimetric and thermal properties in evolutionary perspectives was designed by \citet{Kolokolova2007}. The authors showed a trend that the shorter a comet spends in the vicinity of the Sun, the higher maximum polarisation ($P_{\rm max}$) in the Red domain and the stronger 10 $\mu$m silicate emission feature it shows. Given that $P_{\rm max}$ is typically observed at phase angles of $\alpha_{\rm max}$ $\sim$ 95\degree, however, only a handful of comets have been observed at such high phase angles. Most comets are observed only at $\alpha$ < 50\degree, and their value of $P_{\rm max}$ can only be inferred from extrapolation to $\alpha_{\rm max}$ with considerable uncertainties.
In order to retrieve $P_{\rm max}$ in at least 10 \% uncertainty, observational data taken up to $\alpha$ $\sim$ 70\degree\ are required \citep{Penttila2005}. For these reasons, we felt the necessity to study the correlation of the observed parameters with a modified approach. Instead of representing the polarising properties of a given comet's dust by $P_{\rm max}$, we employ its mean relative deviation from the average trend, $P_{\rm excess}$, as an indicator.

In Section \ref{sec:obsdata}, we describe the datasets used in this study, followed by the results and discussion in Section \ref{sec:res}. In Section \ref{sec:concl}, we present our conclusions.

\section{Data Description \label{sec:obsdata}}

We retrieved the degree of linear polarisation ($P$) of cometary dust from the NASA Planetary Data System (PDS): Small Bodies Node archival data \citep{Kiselev2017}. Given $\sim$2,700 measurements for $\sim$70 comets ranging from $\sim$0.31 to 2.39 $\mu$m, we considered two spectral regions: the Red domain, 0.62--0.73 $\mu$m, in which the number of data points is the largest; and the K domain of 2.00--2.39 $\mu$m.
From the Red domain data, we selected only narrow-band and spectropolarimetric observations to minimise the flux contribution by depolarising gas molecules \citep[e.g.,][]{Kwon2017}. In the K domain and at heliocentric distances of $r_{\rm H}$ $<$ 0.9 au, thermal emission contributes a significant ($\sim$30 \% at most) fraction of the total light received which could lead to lower observed $P$ than would be received from the scattered light only \citep{Oishi1978a}. For comet C/1975 V1 (West), with observations at $r_{\rm H}$ ranging from 0.341 to 0.941 au, we used thermally corrected results provided by \citet{Oishi1978a}. Except for comet 55P/Tempel-Tuttle and one point of C/1995 O1 (Hale-Bopp) having $r_{\rm H}$ of 0.99 au and 0.96 au, respectively, all other observations were made at $r_{\rm H} >$ 1 au; thus, we can exclude the dominance of thermal emission in the polarised signals.

\begin{figure}[!htb]
\centering
\includegraphics[width=9cm]{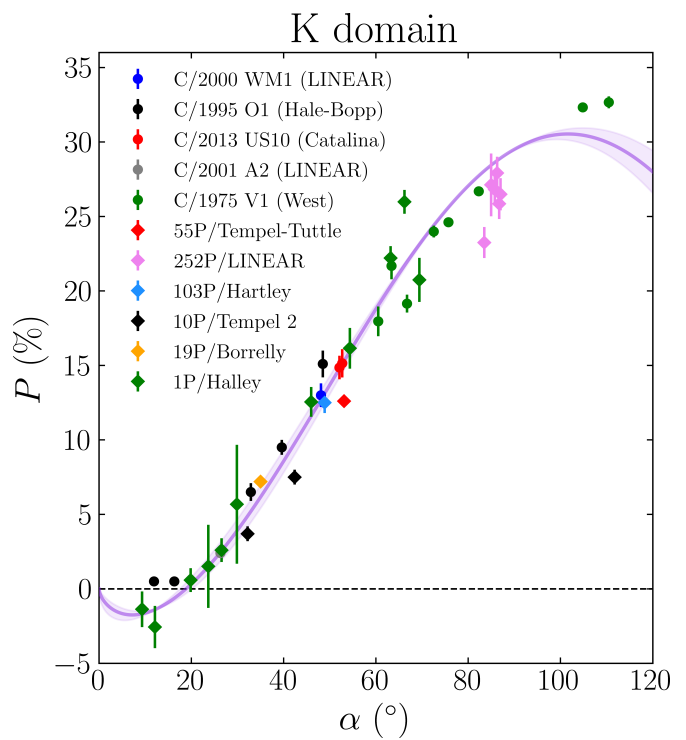}
\caption{Polarization of cometary dust in the K domain as a function of $\alpha$. The solid curve denotes the average $\alpha$ dependence from Eq. \ref{eq:eq1}. The coloured area covers the 3$\sigma$ region of the interpolation. The best fit (minimum $\chi$$^{\rm 2}$) parameters are $b$ = 31.17 $\pm$ 0.48 \%, $c_{\rm 1}$ = 0.62 $\pm$ 0.05, $c_{\rm 2}$ = (3.41 $\pm$ 0.04) $\times$ 10$^{\rm -31}$, and $\alpha_{\rm 0}$ = 18\fdg86 $\pm$ 0\fdg44. Diamonds and circles denote points of short-period comets (including Halley-Types) and long/non-period comets, respectively.}
\label{Fig01}
\vskip-1ex
\end{figure}

\begin{figure}[!t]
\centering
\includegraphics[width=9cm]{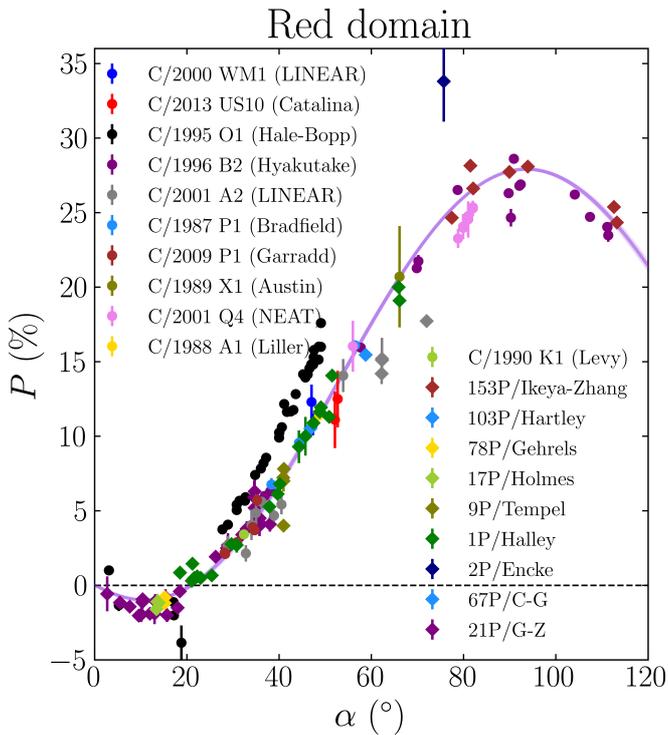}
\caption{Same as Figure \ref{Fig01} but in the Red domain. The best fit parameters are $b$ = 34.85 $\pm$ 0.22 \%, $c_{\rm 1}$ = 1.09 $\pm$ 0.02, $c_{\rm 2}$ = 0.46 $\pm$ 0.01, and $\alpha_{\rm 0}$ = 20\fdg98 $\pm$ 0\fdg24. }
\label{Fig02}
\vskip-1ex
\end{figure}

Figures \ref{Fig01} and \ref{Fig02} show the $P$ distribution of the selected comets as a function of the phase angle ($\alpha$; the angle of the Sun--Comet--Observer) in the K and Red domains, respectively. We fitted the average phase curve in each domain with the empirical trigonometric function of \citet{Lumme1993}:
\begin{equation}
 P(\alpha) = b~\sin^{c_{\rm 1}}(\alpha) \times \cos^{c_{\rm 2}} \bigg(\frac{\alpha}{2}\bigg) \times \sin(\alpha - \alpha_{\rm 0})~,
\label{eq:eq1}
\end{equation}
\noindent where $b$, $c_{\rm 1}$, $c_{\rm 2}$, and $\alpha_{\rm 0}$ are the wavelength-dependent free parameters shaping the curve. We specified the best-fit parameters in the figure captions. 

We next divided the observed values of $P$ by the fitted one at the given $\alpha$. For each comet, we calculated the mean of this ratio, $P_{\rm excess}$, from all available observations weighted by the square of the errors of the data points to investigate the deviation of its $P$ value from the mean behaviour of comets. We used only data from $\alpha$ $>$30\degree\ to avoid the influence of an interference effect of internal waves within the dust particles (i.e., coherent-backscattering effect) that prevails in the negative polarisation branch ($\alpha$ $\lesssim$ 22\degree\ for cometary dust; \citealt{Muinonen2015}). The resultant $P_{\rm excess}$ values are tabulated in Table \ref{t1} with detailed descriptions in Appendix \ref{sec:app1}.

For comets with available polarimetric information as described above, we compiled the flux ratio of the 10 $\mu$m silicate emission to the local continuum ($F_{\rm Si}/F_{\rm cont}$) from the literature. If for a given comet this ratio had been measured during multiple epochs, we use the weighted-mean $F_{\rm Si}/F_{\rm cont}$ of the individual measurements. If no thermal observations were conducted for a comet at a similar epoch with its polarimetric study, we employ a weighted mean of $F_{\rm Si}/F_{\rm cont}$ values taken around the smallest $r_{\rm H}$, i.e., likely around the peak of comet activity. We assigned $F_{\rm Si}/F_{\rm cont}$ = 1.00 $\pm$ 0.15 for comets showing no excess silicate emission.
Consequently, a total of 14 and 9 comets in the Red and K domains, respectively, was utilised to search for a relationship between  polarimetric and thermal properties of the dust particles. Their characteristics and references are summarised in Table {\ref{t1}}, and the ancillary information of the comets used for the analysis (e.g., the aperture size and the observing epoch) is provided in Appendix \ref{sec:app1}.

\section{Results and Discussion \label{sec:res}}

\begin{figure}[!b]
\centering
\includegraphics[width=9cm]{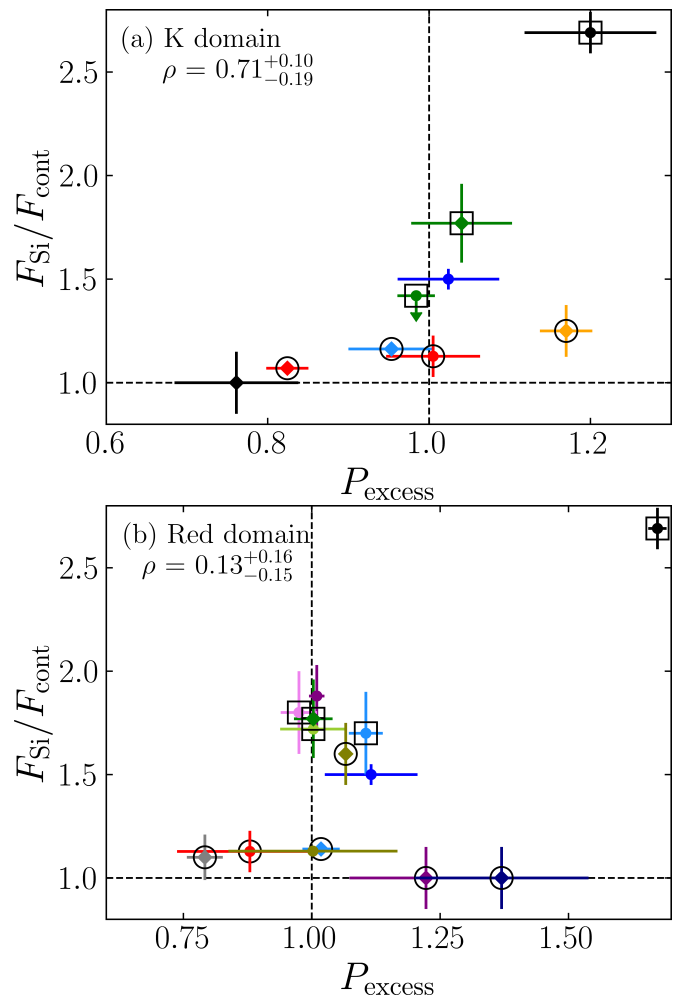}
\caption{Interrelations between $P_{\rm excess}$ and $F_{\rm Si}/F_{\rm cont}$ of cometary dust in the K (a) and Red (b) domains, respectively, with Spearman's rank correlation coefficient ($\rho$) shown upper left in each panel. The meaning of symbols is the same as in Figures \ref{Fig01} and \ref{Fig02}. Black circles and squares around the comet symbols indicate carbon-rich and silicate-rich comets, respectively, as defined in the main text.}
\label{Fig03}
\vskip-1ex
\end{figure}

Figure \ref{Fig03} shows the interrelation between $P_{\rm excess}$ and $F_{\rm Si}/F_{\rm cont}$ of the cometary dust in the K (a) and Red (b) domains. Our method to calculate Spearman's rank correlation coefficient ($\rho$) and its errors is described in Appendix \ref{sec:app2}.
In the K domain, we observe a strong positive correlation between the two quantities ($\rho$ = 0.71$^{+0.10}_{-0.19}$), while they appear nearly uncorrelated in the Red domain ($\rho$ = 0.13$^{+0.16}_{-0.15}$).
  
We put open circles and squares around carbon-rich and silicate-rich comets, respectively, to see whether this metric yields a meaningful classification of the comets. We count a comet as carbon-rich if its silicate-to-amorphous carbon mass ratio (Si/C) is less than unity. This ratio is typically estimated by MIR spectral modelling (see, e.g., \citealt{Swamy1988} and \citealt{Wooden1999}). For 67P/Churyumov-Gerasimenko (67P), we used in-situ results showing the Si/C $\sim$ 0.2  \citep{Bardyn2017}. There are no Si/C values reported for comets C/2000 WM1 (LINEAR), 10P/Tempel 2, C/1989 X1 (Austin), and C/1996 B2 (Hyakutake).

Despite incomplete information, carbon-rich comets in both domains appear to have systematically weaker silicate emission features than silicate-rich comets. However, $P_{\rm excess}$ values of carbon-rich comets are distributed widely horizontally for both wavebands, showing no difference between silicate-rich and carbon-rich comets in polarisation. We conclude from Figure \ref{Fig03} that the dust composition would be a peripheral factor in explaining the observed interdependences of $P_{\rm excess}$ and $F_{\rm Si}/F_{\rm cont}$. 
Moreover, supposing that the scattering cross-section of coma dust is dominated by non-Rayleigh particles \citep{Fulle2000,Kolokolova2004}, we would expect a composition-driven dependence of the two metrics to manifest as an anti-correlation rather than the observed positive correlation. That is because more strongly absorbing (carbon-rich) dust would yield a higher $P$ than transparent dust \citep{Zubko2014,Zubko2016}, and suppress silicate features by enhancing the underlying pseudo-continuum, featureless signal (see, e.g., \citealt{Wooden2002}).

These two considerations lead us to dismiss dust composition as the main parameter behind the $P_{\rm excess}$--$F_{\rm Si}/F_{\rm cont}$ correlation. Instead, we consider the dust size as a main controlling factor. Small particles in the Mie scattering regime (order of 0.1--1 $\mu$m in the NIR) are hotter and show a more robust contrast of 10 $\mu$m silicate emission over the continuum than larger particles \citep{Lisse2004,Sitko2004}. Such small particles also show higher $P$ than larger ones \citep{Kimura2006}; thus, dust size can explain the observed positive correlation of the parameters.

To further explore this hypothesis, we divided the comets into ``gas-rich'' and ``dust-rich'' groups and searched for their distribution in the same two-dimensional parameter space as used in Figure \ref{Fig03}. The criteria for classification are (i) $W$ (the flux ratio of C$_{\rm 2}$ emission bands at 5140 \AA\ to the local continuum at 4845 \AA) values or any similar gas to continuum flux ratio estimates, where dust-rich and gas-rich comets typically show $W$ $<$ 500 and $W$ $>$ 1000 \citep{Swamy2010}; or (ii) if no $W$ for use, the aperture dependence of $P$ in broad-band filters. The latter criterion is based on Figure 1 in \citet{Kolokolova2007} which suggests that in gas-rich comets (according to criterion (i)) the average $P$ across the coma steeply decreases with aperture size, while in dust-rich comets, the aperture-dependence of $P$ is weak. We consider comets with unmeasured $W$ but showing a steep decrease of the average $P$ ($\gtrsim$30 \%) as the radial distance increases out to the order of 10$^{\rm 4}$ km as ``gas-rich.''

\begin{figure}[!t]
\centering
\includegraphics[width=9cm]{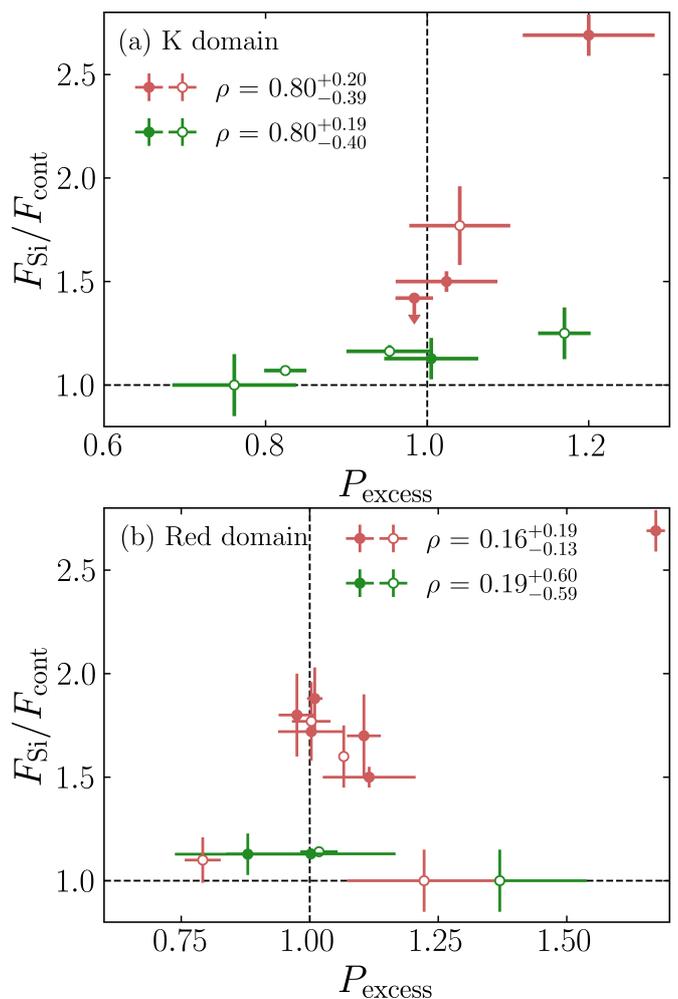}
\caption{Same as Figure \ref{Fig04} but comets are classified in ``dust-rich'' (red), and ``gas-rich'' (green) comets in the K (a) and Red (b) domains, respectively, with $\rho$ values provided upper left for each group.  The classification criteria are given in the text. Open and filled circles denote points of short-period comets (including Halley-Types) and long/non-period comets, respectively.
}
\label{Fig04}
\vskip-1ex
\end{figure}

Figure \ref{Fig04} shows the resultant distribution of the two groups in both spectral domains. In K-band, although the ``gas-rich'' comets are located below the ``dust-rich'' comets, within each group we find a strong positive correlation between $P_{\rm excess}$ and $F_{\rm Si}/F_{\rm cont}$. The correlation coefficient is comparable for both groups ($\rho$ $\sim$ 0.80) and stronger than for the dataset as a whole in Figure \ref{Fig03}. Also in the Red domain, the gas-rich comets are on average located lower than the dust-rich comets, but no discernible correlation of the two parameters is shown.

In the following, we aim to gain insight into which dust property is mainly driving the correlation in K-band observed in Figures \ref{Fig03} and \ref{Fig04}. As mentioned above, gas-rich comets observed at 0.5--0.9 $\mu$m have in common that they frequently show a steep decrease of the observed average $P$ over the outer coma, along with the increase in the flux ratio of gas molecules to the total signal. This contrasts with the behaviour of dust-rich comets that show almost constant radial $P$ dependence within $<$10\%\ \citep{Kolokolova2007}.
When $P$ is measured in broad-band filters with the spatial resolution of thousands of km typical for ground-based telescopes, it thus tends to be lower in ``gas-rich'' comets than in ``dust-rich'' comets. However, if measured in narrow-band filters or with spectropolarimetry, the ``gas-rich'' comets also show their intrinsic dust $P$ values because depolarised light in emission lines is excluded from the measurement. The resulting aperture-averaged $P$ is not as low as observed in broad-band and is nearly constant across the coma \citep{Jockers2005,Kwon2017,Kwon2018}. 

\citet{Kolokolova2007} suggest that such aperture-size dependence of $P$ could be related to the radial distribution of the coma dust, which in turn could reflect the dust porosity. In this interpretation, ``gas-rich'' comets have relatively heavy, compact dust particles lingering around the nucleus and thus show higher gas contamination in the wide aperture broad-band polarimetric data. In contrast, more porous dust in ``dust-rich'' comets has a larger cross-section-to-mass ratio that reaches farther cometocentric distance under the same condition yielding more or less constant $P$ radial variations. For Figure \ref{Fig04}, this would mean that the green symbols refer to comets with predominantly ``compact'' dust, while red symbols indicate more porous dust. This interpretation is also consonant with the dust modelling results showing that highly fluffy particles (BCCA) can retain strong silicate emission features even for 100 $\mu$m-sized dust, while more compact ones (BPCA) show featureless MIR spectra \citep{Kimura2009}\footnote{\citet{Chornaya2020} recently suggested that even half-mm-sized compact olivine particles show the 10 $\mu$m silicate features. However, the facts that (i) \citet{Chornaya2020} measure the reststrahlen-band features in reflectance spectra (Section 9 in \citealt{Bohren1983}), not emission features in thermal spectra, and (ii) such large, transparent particles cannot explain the observed temperature of cometary dust (e.g., \citealt{Wooden2002,Hanner2004}) make it hard to compare their results to this study directly.}. Furthermore, given some overlap between the ``carbon-rich'' (Figure \ref{Fig03}) and ``gas-rich'' (Figure \ref{Fig04}) comets, compact dust in the ``gas-rich'' comets might produce weaker silicate features that could be attributed to smaller Si/C ratios and hence lead to the comets being classified as carbon-rich. Hence, both the distinction between ``gas-rich'' and ``dust-rich'' comets and that between ``carbon-rich'' and ``silicate-rich'' could in truth be the one between ``porous'' and ``compact'' dust.

In-situ ESA/Rosetta observations have revealed the hierarchical nature of dust of comet 67P from m-scale boulders down to the sub-$\mu$m monomers \citep{Bentley2016}. Observations in 67P's innermost coma showed that the ejected dust is barely present as individual grains but bound in aggregates (or agglomerates), which can be classified by different mechanical strengths depending on their bulk porosity \citep{Fulle2016,Guttler2019}. The number density of fluffy and compact particles varies with 67P's season  \citep{Longobardo2020}: dust ejected from more evolved, smooth terrains shows lower porosity, whereas dust from fresh, rough terrains shows higher porosity. Hence, porosity may be indicative of the degree of processing experienced by the dust, which is also in line with the conclusions of \citet{Kolokolova2007}.

Fluffy dust particles with a tensile strength of $<$10$^{\rm 5}$ N m$^{\rm -2}$ \citep{Mendis1991} and a high charge-to-mass ratio \citep{Fulle2015} are more vulnerable to disintegration due to either electrostatic fragmentation, ice sublimation and/or radiative torques \citep{Herranen2020} than low porosity dust. In the end, the high ratio of fluffy versus compact dust particles naturally leads to an increase in the small ($\lesssim$$\mu$m-sized) particles in the size distribution of the coma dust as it moves out from the nucleus. This results in a higher $P_{\rm excess}$ and $F_{\rm Si}/F_{\rm cont}$ for observations with a thousand-km-sized aperture.
Consequently, the dust size, related to the porosity of dust aggregates, would be the determinant to make the two parameters correlated positively. The strong correlation in the K domain would demonstrate that different comets have different average dust porosity, which might be an evolutionary outcome with compact particles being more strongly processed.

Why are then such $P_{\rm excess}$--$F_{\rm Si}/F_{\rm cont}$ correlations absent in the Red domain? It might be explained by the different scattering scales. At a wavelength of $\sim$0.6 $\mu$m, monomers in aggregates have more or less similar dimensions with the incident wavelet \citep{Guttler2019}, such that their individual characteristics define the scattering to a large extent. In contrast, the single wavelet of $\sim$2.2 $\mu$m can cover multiple monomers, enhancing mutual electromagnetic interaction ($\propto$$d^{\rm -3}$, where $d$ is the distance between two neighbouring dipole-like monomers; \citealt{Jackson1962}), which makes observations more pertinent to the way of dust organisation in aggregates, i.e., bulk effects \citep{Kolokolova2010, Kwon2019}. The apparent lack of a correlation in the Red domain might thus tell us that the monomers' inherent characteristics would be independent of porosity. Panel b of Figure \ref{Fig04} also shows that $P_{\rm excess}$ is not correlated with the dynamic group of the comets. Such a broad similarity of monomers' traits between different dynamic groups might further support the similar origin of short- and long-period comets, as suggested by recent dynamical studies \citep{Morbidelli2015}.

We should emphasise that it is premature to make statistically significant conclusions from this study due to the small datasets. The outliers in the Red domain, particularly 67P and 2P/Encke (purple and navy diamonds, respectively, in Figure \ref{Fig03}), also have no corresponding K-domain polarimetric data, making it challenging to assess whether the observed trend in the K domain is unbiased. Further coordinated studies of polarimetry and MIR spectroscopy are needed to draw a more comprehensive conclusion.

\section{Conclusions \label{sec:concl}}

We present a new analysis searching for the possible correlation between polarimetric and thermal silicate emission properties of cometary dust. We estimated the relative excess of comets' polarisation status about the average trend at the given phase angle, parameterised as $P_{\rm excess}$. The 10 $\mu$m silicate emission excess over the continuum $F_{\rm Si}/F_{\rm cont}$ was then used to examine the relationship with $P_{\rm excess}$.

The K domain $P_{\rm excess}$ and $F_{\rm Si}/F_{\rm cont}$ are positively correlated for the comets as a whole and also for gas-rich and dust-rich comets individually, enabling us to claim that (i) the dust composition is  secondary in interpreting the correlated observations; and (ii) the dust size and porosity would be key factors instead.

We propose that the current classifications in ``gas-rich'' and ``dust-rich'' on the one hand, and in ``carbon-rich'' and ``Si-rich'' on the other might in reality be both related to the difference between ``low'' and ``high'' porosity, especially as the two classification schemes have considerable overlap in our data set.

With Rosetta results indicating that less porous dust might be more strongly processed, we suggest that the porosity inferred from $F_{\rm Si}/F_{\rm cont}$ might help to diagnose the compaction of a comet's dust as a consequence of comet evolution.

In contrast, the Red domain $P_{\rm excess}$ seems uncorrelated with $F_{\rm Si}/F_{\rm cont}$. Accordingly, we conclude that optical polarimetry is mainly sensitive to the properties of monomers inside an aggregate, while the NIR polarimetry is useful to diagnose the porosity of the dust aggregates. The distributions in the Red domain might further imply the broad similarity of monomers and hence in the origin of short-period and non/long-period comets.

Our results also suggest that the current MIR spectral models of cometary dust, which search for a best fit by simultaneously considering various dust properties (e.g., the size distribution, composition, and porosity; \citealt{Wooden2004,Harker2018,Woodward2021}, should prioritise the structural parameters over the composition. We recommend coordinated studies in polarimetry and MIR spectroscopy in order to shed further light on the processes shaping cometary dust particles in our planetary system.
\\

\begin{acknowledgements}
Y.G.K. gratefully acknowledges the support of the Alexander von Humboldt Foundation. J.A. acknowledges funding by the Volkswagen Foundation. J.A. and J.M. acknowledge funding from the European Union’s Horizon 2020 research and innovation programme under grant agreement No 757390 CAstRA.

\end{acknowledgements}


\begin{appendix}

\section{Ancillary Information for Comets Used for Analyses\label{sec:app1}}

\begin{figure}[!b]
\centering
\includegraphics[width=9cm]{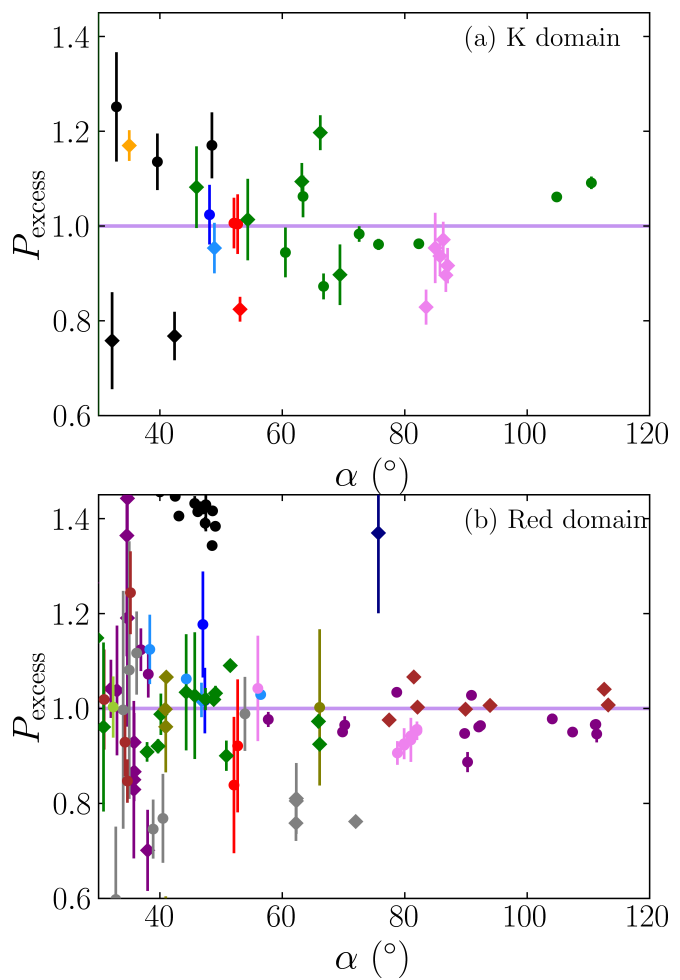}
\caption{$P_{\rm excess}$ of comets in the K and Red domains as a function of $\alpha$. The $P_{\rm excess}$ = 1 line indicates the interpolated average $\alpha$ dependence derived from Eq. \ref{eq:eq1}. }
\label{Fig05}
\vskip-1ex
\end{figure}

The polarimetric and MIR observations for most of the comets were made in approximately the same periods. If a comet was observed in multiple epochs and the period span of its polarimetric observations overlaps the span of its MIR observations, we considered all the overlapped data points. For five comets (C/2013 US10 (Catalina), 19P/Borrelly, 55P/Tempel-Tuttle, C/1990 K1 (Levy), and 2P/Encke), we found no common periods between polarimetric and MIR observations. However, based on the following reasons, we expect two observing modes to observe nearly similar environments of the comets. For C/2013 US10 (Catalina), both observations were made post-perihelion in which the heliocentric difference ($\Delta r_{\rm H}$) is $\sim$0.27 au. The MIR observations for 19P/Borrelly were made one apparition before the polarimetric observations, but at the same $r_{\rm H}$. The MIR observations of 55P/Tempel-Tuttle were conducted nine days before the polarimetric observations, but $\Delta r_{\rm H}$ is less than 0.03 au. The MIR observations of C/2001 Q4 (NEAT) were conducted one week before the polarimetric observations, but $\Delta r_{\rm H}$ is less than 0.007 au. The polarimetric observations of 2P/Encke were made pre-perihelion, while its MIR observations were made at and post-perihelion, with $\Delta r_{\rm H}$ $\sim$ 0.16 au. Nonetheless, since 2P/Encke consistently showed no silicate emission features over the local continuum throughout the MIR observations \citep{Gehrz1989}, the usage of $F_{\rm Si}/F_{\rm cont}$ = 1 would not be problematic. Although it turns out that the intensity of the silicate emission of a comet could vary within a few days (e.g., \citealt{Mason2001}), we expect that the usage of the weighted mean of multiple data points taken at similar epochs of the two observing modes would represent a comet's general behaviour relative to the average trend of other comets.

The difference in the aperture size applied between the polarimetric and MIR observations was also examined. Most of the comets were sampled by the similar aperture size in the two observing modes, except for the  five comets (C/2001 WM1 (LINEAR), 55P/Tempel-Tuttle, C/1990 K1 (Levy), 21P/Giacobini-Zinner, and 9P/Tempel 1) having different aperture sizes but in the same order. The first four comets were observed in two observing modes with a few and a few tens of thousand km-sized apertures at $r_{\rm H}$ $\gtrsim$ 1 au. Their data points thus represent the dust properties in the dust tail/trail coma region, free from any significant collisions and ice sublimations \citep{Marschall2020}. Together with the results of \citet{Bonev2008} showing that the dust radial profile reaches an equilibrium at $\sim$1,000 km from the nucleus at $r_{\rm H}$ $\sim$ 1 au, the aperture differences in such an order might sample more or less similar coma dust. For comet 9P/Tempel 1 right after the Deep Impact experiment, on the other hand, both polarimetric and MIR observations applied aperture sizes less than 1,000 km at $r_{\rm H}$ $\sim$ 1.5 au. Since both MIR and polarimetric observations detected impact-induced changes almost simultaneously (e.g., \citealt{Furusho2007,Harrington2007,Harker2005,Kuppers2005,Harker2007}), we expect that the coma dust properties observed by the two observing modes would not be absurdly different. The only comet having an irreconcilable difference in the aperture sizes is C/2001 Q4 (NEAT). Its polarimetric data cover $\sim$7,100 to 20,000 km cometocentric distance, while the MIR data were extracted $\sim$760 km from the centre. Although the usage of the weighted mean of its multiple data points places the comet around the centre in the $P_{\rm excess}$ and $F_{\rm Si}/F_{\rm cont}$ plane (pink circle in Figure \ref{Fig03}b), consistent with the previous study showing its silicate spectrum similar to those of other long-period comets \citep{Wooden2004}, the interpretation of this comet's data points should be made with caution due to the systematic difference of the covered coma scale.

Despite the limitations mentioned above, we expect that the usage of the weighted mean of the multi-epoch data points across the wide range of the phase angles could mitigate the systematic differences for some comets and enable us to examine the physical and/or compositional status of a comet relative to the majority of comets. The observed $P$ values in Figures \ref{Fig01} and \ref{Fig02} are divided by the best-fit average at the given $\alpha$, as shown in Figure \ref{Fig05}. The symbols are identical to the ones for Figures \ref{Fig01} and \ref{Fig02}. Since a single comet taken at different $\alpha$ frequently shows fluctuations of $P$ values around the average trend line, we took a weighted mean for each comet, parameterised as $P_{\rm excess}$, to address the general polarimetric status of a comet with regard to the majority of comets:
\begin{equation}
P_{\rm excess} = \frac{\sum_{i=1}^{n} w_{i}~P_{\rm excess, {\it i}}}{w_{i}}
~,
\label{eq:eq2}
\end{equation}
\noindent where $n$ is the number of $P$ data points at $\alpha$ > 30\degree\ for a single comet, $P_{\rm excess, {\it i}}$ is the $P_{\rm excess}$ of the $i^{\rm th}$ point of the comet, and $w_{i}$ is the weight of the $i^{\rm th}$ point of the comet, which is 1/$\sqrt{\sum_{i=1}^{n} (\sigma P_{\rm excess, {\it i}})^{\rm -2}}$ where $\sigma P_{\rm excess, {\it i}}$ means the error on $P_{\rm excess, {\it i}}$. The error of $P_{\rm excess}$ is the standard deviation of $P_{\rm excess, {\it i}}$ divided by $\sqrt{n}$. 
Likewise, the weighted mean of $F_{\rm Si}/F_{\rm cont}$ (when multiple epochs are considered for a comet) was retrieved under the same scheme with Eq. \ref{eq:eq2}, instead of weighted by 1/$\sqrt{\sum_{i=1}^{n} (\sigma(F_{\rm Si}/F_{\rm cont})_{i})^{\rm -2}}$ where $\sigma(F_{\rm Si}/F_{\rm cont})_{i}$ is the error on $F_{\rm Si}/F_{\rm cont}$ adopted from the references. For comets showing no excess of the 10 $\mu$m silicate feature (comets 10P/Tempel 2, 2P/Encke and 67P), we found no available information about neither the signal-to-noise ratios of the spectra nor the uncertainties of the fitting models of the local continuum, except for the 10--15 \% uncertainty in the temperature correction used for the underlying continuum fitting of 10P/Tempel 2 \citep{Tokunaga1992}. Hence, we assigned 15 \% error bars for the comets having no silicate emission features.

Table {\ref{t1}} summarises the description of the datasets used for drawing Figures \ref{Fig03} and \ref{Fig04}. Numbers in the last column show references for the $F_{\rm Si}/F_{\rm cont}$, carbon abundance, and the criteria for classifications of gas-rich versus dust-rich comets. 
Polarimetric data are quoted from the NASA PDS database \citep{Kiselev2017} and subsequent studies of \citet{Kwon2017}, \citet{Kwon2018}, and \citet{Kwon2019}. 

\begin{table*}[!h]
\centering
\caption{Observational characteristics of comets used in Figures \ref{Fig03} and \ref{Fig04} and their references}
\vskip-1ex
\begin{tabular}{l|cccccc|c}
\toprule
\toprule
Comet name & Domain & $P_{\rm r}/P_{\rm avg}$ & $F_{\rm Si}/F_{\rm cont}$ & Obs. Type & C. abundance & Dyn. Type & Reference\\
\hline
\multirow{2}{*}{C/1995 O1 (Hale-Bopp)} & Red & 1.673 $\pm$ 0.018 & \multirow{2}{*}{2.690 $\pm$ 0.100} & \multirow{2}{*}{dust-rich} &  \multirow{2}{*}{low} & \multirow{2}{*}{LPC}  & \multirow{2}{*}{(1)(2)(3)}\\
 & K & 1.200 $\pm$ 0.082 &  &  &  &  & \\
\multirow{2}{*}{C/2013 US10 (Catalina)} & Red & 0.880 $\pm$ 0.142 & \multirow{2}{*}{1.128 $\pm$ 0.100} & \multirow{2}{*}{gas-rich} & \multirow{2}{*}{high} & \multirow{2}{*}{NPC} & \multirow{2}{*}{(4)(5)}\\
 & K & 1.005 $\pm$ 0.058 &  &  &  &  & \\
\multirow{2}{*}{C/2000 WM1 (LINEAR)} & Red & 1.116 $\pm$ 0.090 & \multirow{2}{*}{1.500 $\pm$ 0.050} & \multirow{2}{*}{dust-rich} & \multirow{2}{*}{$-$} & \multirow{2}{*}{NPC} & \multirow{2}{*}{(6)(7)}\\
 & K & 1.024 $\pm$ 0.063 &  &  &  &  &  \\
\multirow{2}{*}{1P/Halley} & Red & 1.003 $\pm$ 0.038 & \multirow{2}{*}{1.770 $\pm$ 0.190} & \multirow{2}{*}{dust-rich} & \multirow{2}{*}{low} & \multirow{2}{*}{HTC} & \multirow{2}{*}{(8)(9)(10)}\\
 & K & 1.040 $\pm$ 0.062 &  &  &  &  &  \\
\multirow{2}{*}{103P/Hartley 2} & Red & 1.018 $\pm$ 0.036 & \multirow{2}{*}{1.163 $\pm$ 0.030} & \multirow{2}{*}{gas-rich} &  \multirow{2}{*}{high} & \multirow{2}{*}{JFC} & \multirow{2}{*}{(11)(12)} \\
 & K & 0.953 $\pm$ 0.053 &  &  &  &  &  \\
10P/Tempel 2 & K & 0.761 $\pm$ 0.079 & 1.000 $\pm$ 0.150 & gas-rich & $-$ & JFC & (13)(14)(15) \\
19P/Borrelly & K & 1.170 $\pm$ 0.032 & 1.250 $\pm$ 0.125 & gas-rich & high & JFC & (16)(17) \\
55P/Tempel-Tuttle & K & 0.824 $\pm$ 0.026 & 1.070 $\pm$ 0.020 & gas-rich & high & HTC & (18)(19) \\
C/1975 V1 (West) & K &  0.984 $\pm$ 0.023 &  1.420 $\pm$ 0.080 & dust-rich & low & NPC & (20)(21)\\
C/1987 P1 (Bradfield) & Red & 1.105 $\pm$ 0.033 & 1.700 $\pm$ 0.200 & dust-rich & low & LPC & (9)(22)(23)\\
C/1989 X1 (Austin) & Red & 1.002 $\pm$ 0.165 & 1.130 $\pm$ 0.020 & gas-rich & $-$ & HTC & (19)(24) \\
C/1990 K1 (Levy) & Red & 1.003 $\pm$ 0.065 & 1.720 $\pm$ 0.040 & dust-rich & low & NPC & (9)(25)(26) \\
C/1996 B2 (Hyakutake) & Red & 1.010 $\pm$ 0.015 & 1.880 $\pm$ 0.150 & dust-rich & $-$ & LPC & (27)(28)\\
C/2001 Q4 (NEAT) & Red & 0.975 $\pm$ 0.035 & 1.800 $\pm$ 0.200 & dust-rich & low & NPC & (29)(30)\\
2P/Encke & Red & 1.370 $\pm$ 0.169 & 1.000 $\pm$ 0.150 & gas-rich & high & JFC & (31)(32) \\
9P/Tempel 1$^\dagger$ & Red & 1.066 $\pm$ 0.003 & 1.600 $\pm$ 0.150 & dust-rich & high & JFC & (33)(34)(35) \\
21P/Giacobini-Zinner & Red & 0.792 $\pm$ 0.035 & 1.100 $\pm$ 0.110 & dust-rich & high & JFC & (9)(36)(37)\\
67P/Churuymov-Gerasimenko & Red & 1.223 $\pm$ 0.149 & 1.000 $\pm$ 0.150 & dust-rich & high & JFC & (38)(39)\\
\hline
\bottomrule
\end{tabular}
\tablefoot{Top headers: $P_{\rm r}/P_{\rm avg}$, the relative excess of the polarisation degree with regard to the average value at the given phase angle ($\alpha$); $F_{\rm Si}/F_{\rm cont}$, the relative excess of the 10 $
\mu$m silicate emission feature to the {\bf local} continuum. Numbered references indicate (1) \citet{Mason2001}; (2) \citet{Harker2002}; (3) \citet{Kiselev1997}; (4) \citet{Kwon2017}; (5) \citet{Woodward2021}; (6) \citet{Joshi2003}; (7) \citet{Kelley2004}; (8) \citet{Hanner1987}; (9) \citet{Chernova1993}; (10) \citet{Wooden2002}; (11) \citet{Lara2011}; (12) \citet{Harker2018}; (13) \citet{Tokunaga1992}; (14) \citet{Lynch1995}; (15) \citet{Paganini2012}; (16) \citet{Hanner1996}; (17) \citet{Szabo2002}; (18) \citet{Watanabe2001}; (19) \citet{Sitko2004}; (20) \citet{Oishi1978a}; (21) \citet{Oishi1978b}; (22) \citet{Hanner1990}; (23) \citet{Colangeli1996}; (24) \citet{Joshi1992}; (25) \citet{Lynch1992}; (26) \citet{Lisse1998}; (27) \citet{Kiselev1998}; (28) \citet{Mason1998}; (29) \citet{Wooden2004}; (30) \citet{Ganesh2009}; (31) \citet{Gehrz1989}; (32) \citet{Kwon2018}; (33) \citet{Harker2005}; (34) \citet{Kuppers2005}; (35) \citet{Harker2007}; (36) \citet{Hanner1992}; (37) \citet{Ootsubo2020}; (38) \citet{Hanner1985}; and (39) \citet{Bardyn2017}. 
LPC, NPC, HTC, JFC are the abbreviations of Long-Period Comets, Non-Period Comets, Halley-Type Comets, and Jupiter-Family Comets, respectively. 
\\
$^\dagger$ Right after the Deep Impact experiment}
\label{t1}
\vskip-1ex
\end{table*}

\section{Statistical Aspects\label{sec:app2}}

\begin{figure}[!htb]
\centering
\includegraphics[width=9cm]{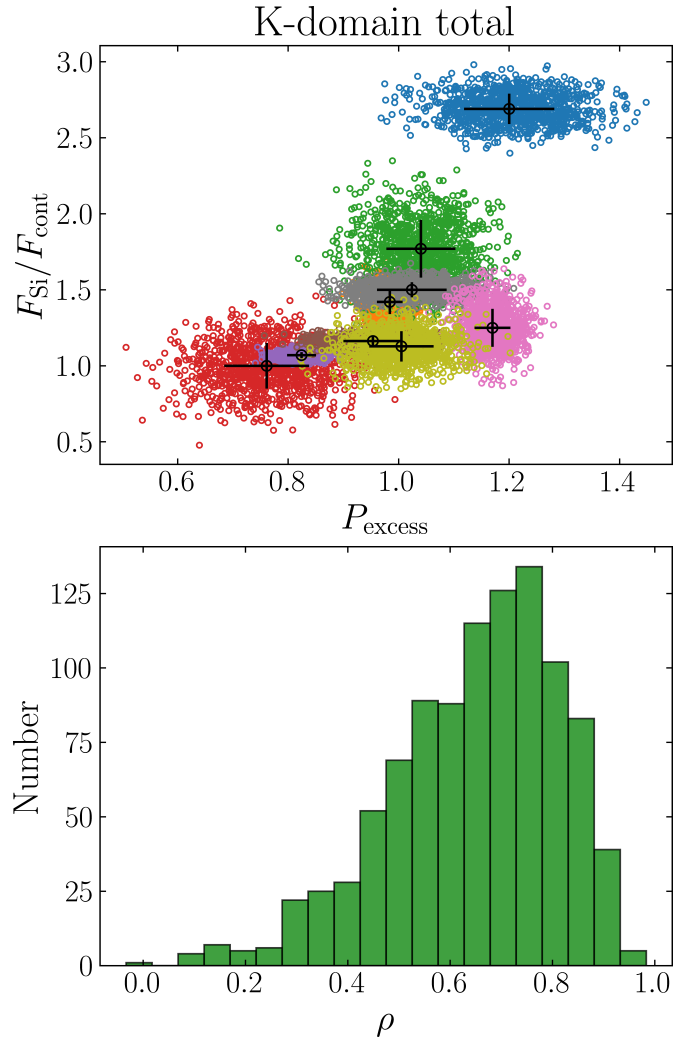}
\caption{Number distribution of clones randomly generated in the Gaussian distribution whose standard deviation corresponds to the error bars of the two parameters of all comets in the K domain in Figure \ref{Fig03}a (a) and histogram of the resultant Spearman's rank correlation coefficient $\rho$ of 1,000 new samples retrieved from Monte Carlo simulation. In panel (a), colours do not correspond with that in Figure \ref{Fig03}. In panel (b), the median of the distribution function with the 68.2 \% interval around the median value is 0.71$^{\rm +0.10}_{\rm -0.19}$ and its mode is 0.76. }
\label{Fig06}
\vskip-1ex
\end{figure}

\begin{figure}[!htb]
\centering
\includegraphics[width=9cm]{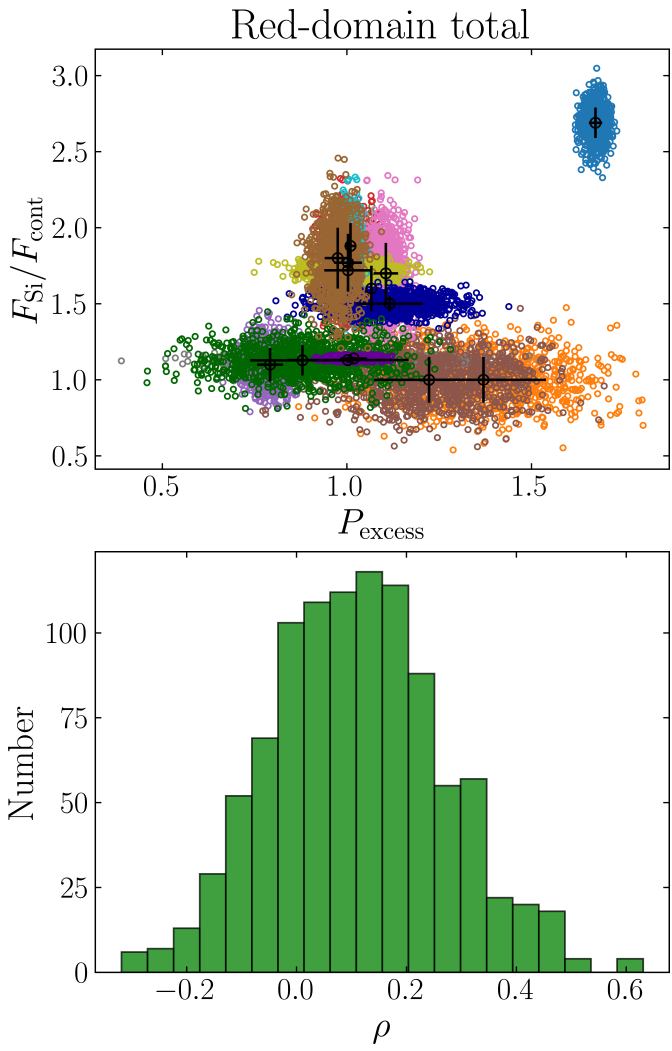}
\caption{Same as Figure \ref{Fig06} but for all comets in the Red domain in Figure \ref{Fig03}b. The median of the distribution function with the 68.2 \% interval around the median value is 0.13$^{\rm +0.16}_{\rm -0.15}$ and its mode is 0.13. }
\label{Fig07}
\vskip-1ex
\end{figure}

\begin{figure}[!htb]
\centering
\includegraphics[width=9cm]{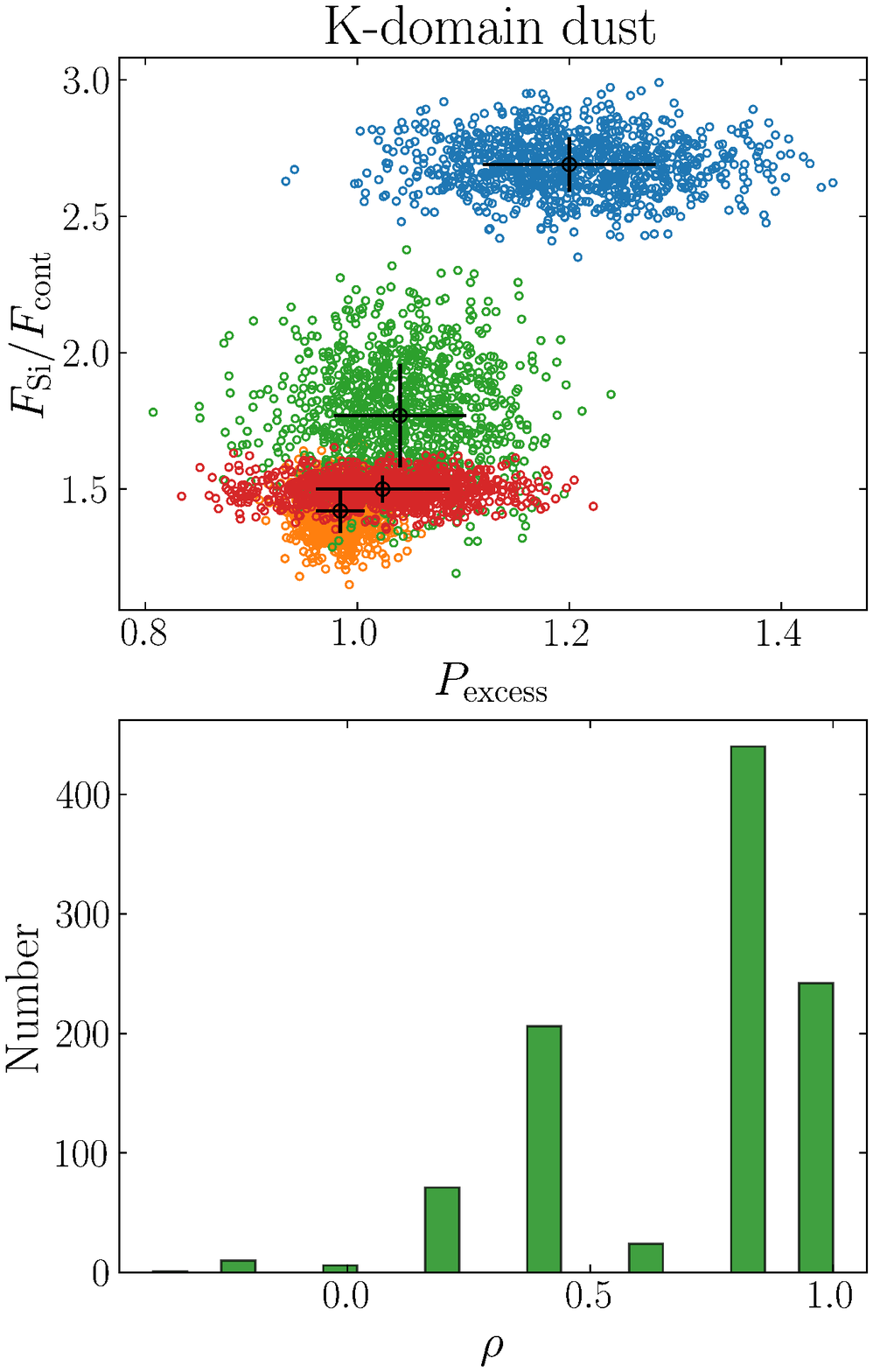}
\caption{Same as Figure \ref{Fig06} but for the dust-rich in the K domain in Figure \ref{Fig04}a. The median of the distribution function with the 68.2 \% interval around the median value is 0.80$^{\rm +0.20}_{\rm -0.39}$ and its mode is 0.83. }
\label{Fig08}
\vskip-1ex
\end{figure}

\begin{figure}[!htb]
\centering
\includegraphics[width=9cm]{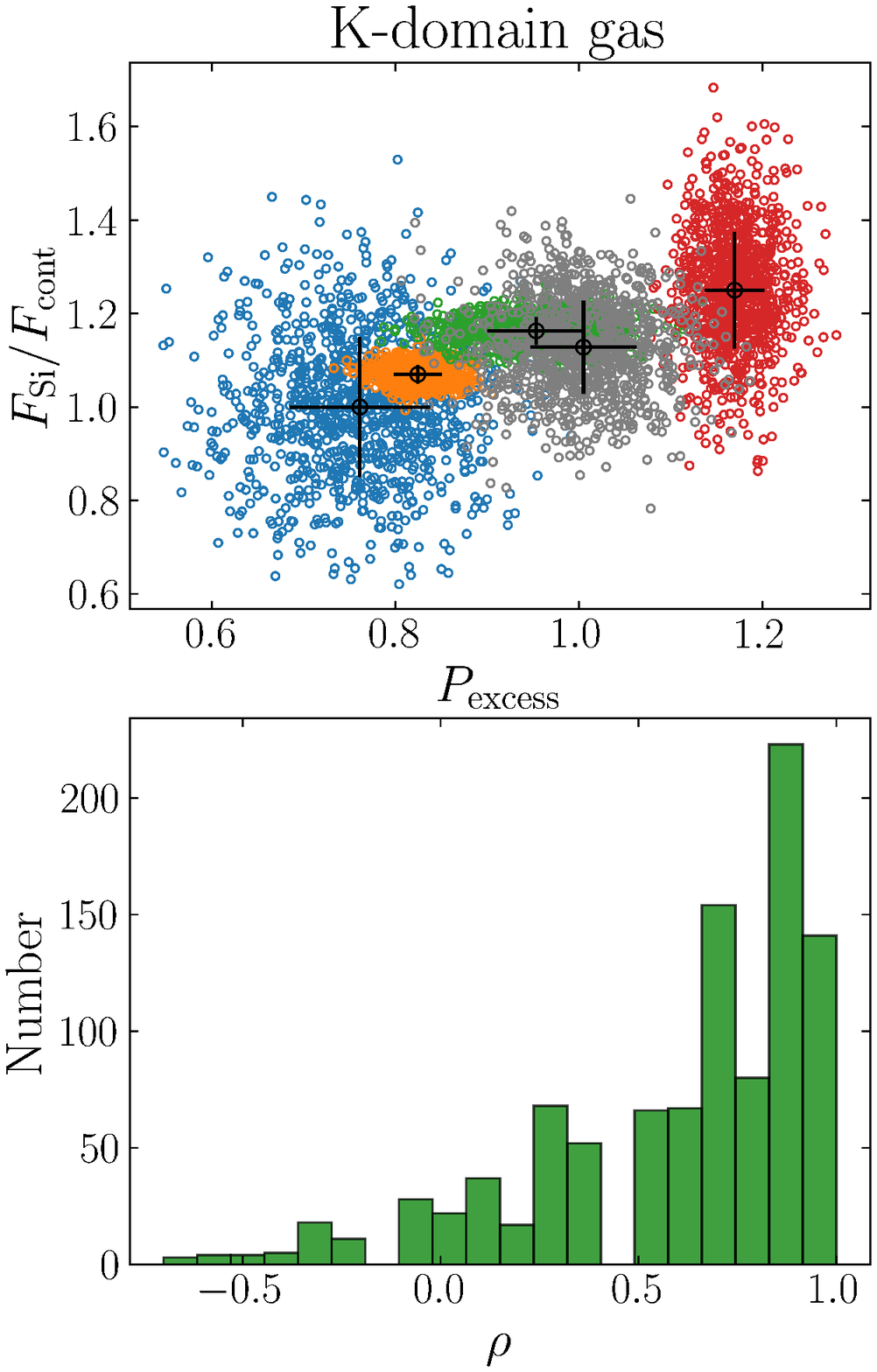}
\caption{Same as Figure \ref{Fig06} but for the gas-rich in the K domain in Figure \ref{Fig04}a. The median of the distribution function with the 68.2 \% interval around the median value is 0.80$^{\rm +0.19}_{\rm -0.40}$ and its mode is 0.87. }
\label{Fig09}
\vskip-1ex
\end{figure}

\begin{figure}[!htb]
\centering
\includegraphics[width=9cm]{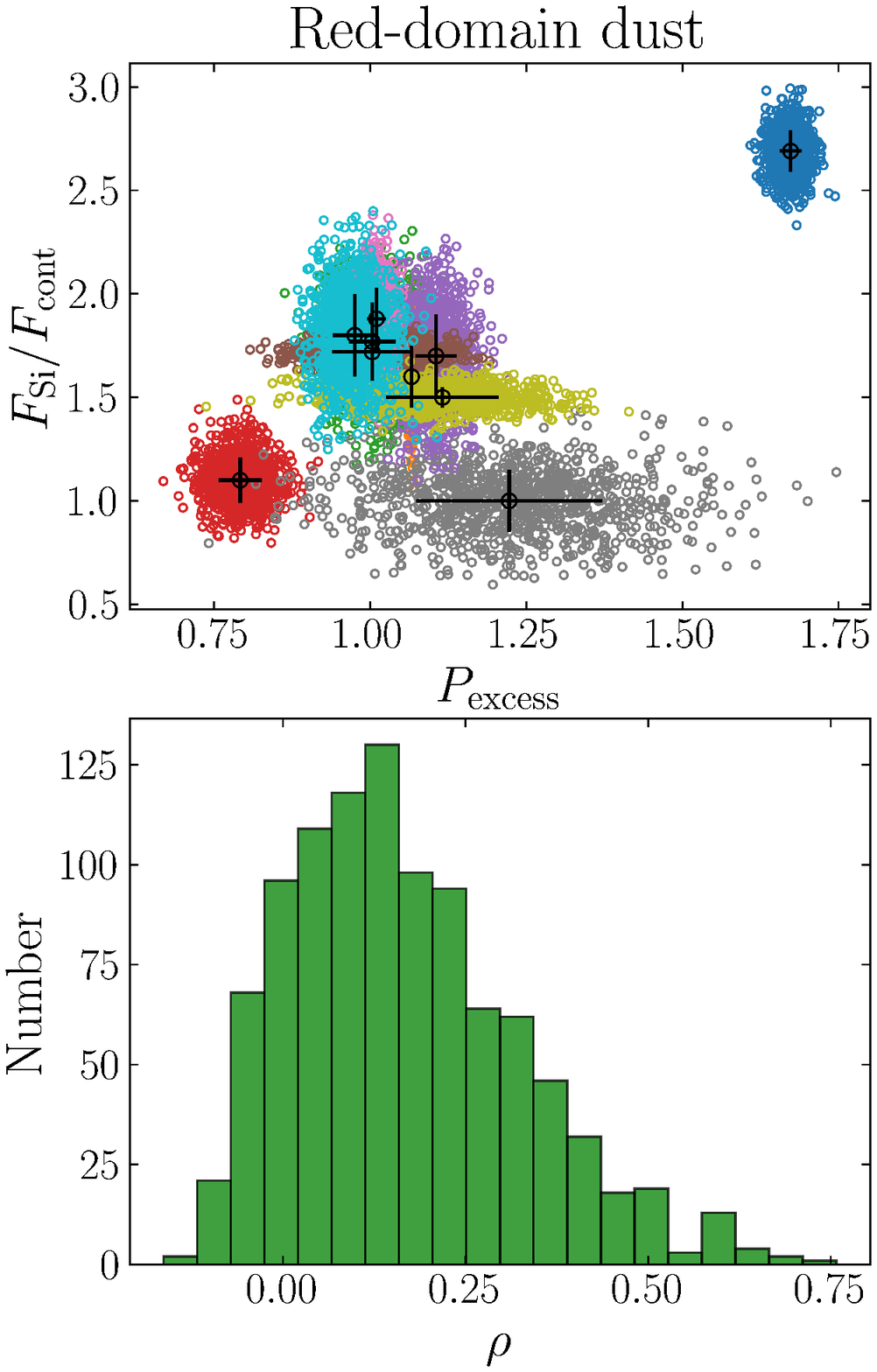}
\caption{Same as Figure \ref{Fig06} but for the dust-rich in the Red domain in Figure \ref{Fig04}b. The median of the distribution function with the 68.2 \% interval around the median value is 0.16$^{\rm +0.19}_{\rm -0.13}$ and its mode is 0.14. }
\label{Fig10}
\vskip-1ex
\end{figure}

\begin{figure}[!htb]
\centering
\includegraphics[width=9cm]{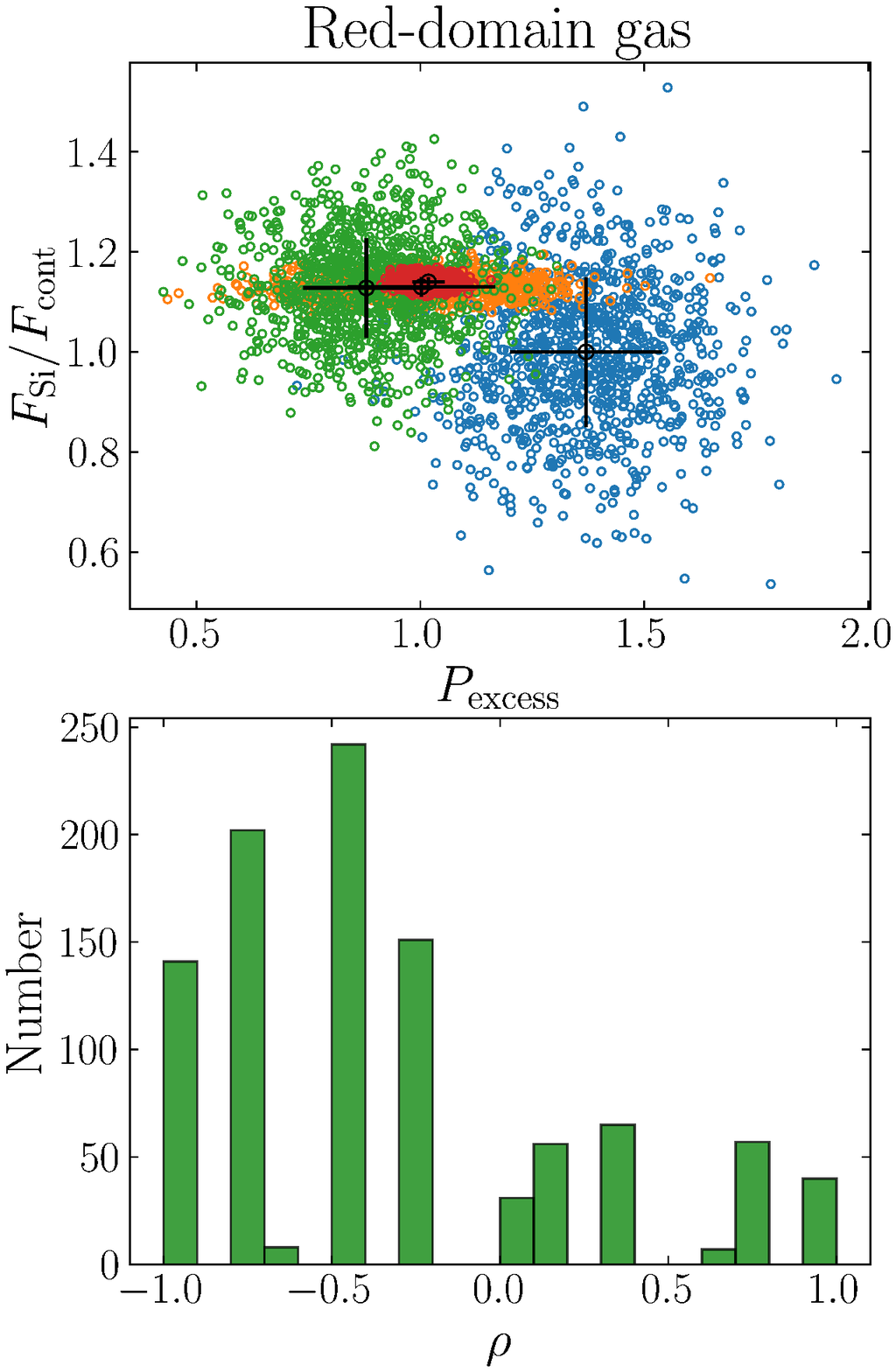}
\caption{Same as Figure \ref{Fig06} but for the gas-rich in the Red domain in Figure \ref{Fig04}b. The median of the distribution function with the 68.2 \% interval around the median value is 0.19$^{\rm +0.60}_{\rm -0.59}$ and its mode is $-$0.45. }
\label{Fig11}
\vskip-1ex
\end{figure}

In order to check for correlations between the two metrics we defined in Section \ref{sec:obsdata} and trends therein, we retrieved the Spearman's rank correlation coefficient, $\rho$. Since the Spearman test does not provide the error bar on the coefficient, we carried out Monte-Carlo simulation to derive more reliable $\rho$ values. We first generated clones of data points randomly distributed in the Gaussian distribution, the standard deviation of which corresponds to the error bars of the two metrics (i.e., the errors of the weighted-mean $P_{\rm excess}$ and $F_{\rm Si}/F_{\rm cont}$ values in Appendix \ref{sec:app1}). Monte-Carlo simulation then retrieved the most probable $\rho$ values by sampling 1,000 synthetic datasets of ($P_{\rm excess}$, $F_{\rm Si}/F_{\rm cont}$), yielding a distribution of $\rho$ with a finite width. The resultant $\rho$ distributions are non-gaussian, such that we adopted the median of the Monte-Carlo results as the nominal value, and the 68.2 \% (1-sigma) interval around the median value as the error bars, as \citet{SantosSanz2012}. Figures \ref{Fig06} and \ref{Fig07} show the distribution of clones and the $\rho$ distribution from the Monte-Carlo simulation for the comets in panels (a) and (b) in Figure \ref{Fig03}, respectively. The nominal $\rho$ value, its error bars, and the mode value are provided in the caption of the figures. The results for the dust-rich and gas-rich comets in the K domain are shown in Figures \ref{Fig08} and \ref{Fig09}, while those in the Red domain are shown in Figures \ref{Fig10} and \ref{Fig11}.


\end{appendix}

\end{document}